\documentstyle[12pt,a4]{article}
\title{Periodic instantons and the loop group}
\author{Paul Norbury\\ norbs@ms.unimelb.edu.au}
\date{}
\newtheorem{lemma}{Lemma}[section]
\newtheorem{cor}[lemma]{Corollary}
\newtheorem{prop}[lemma]{Proposition}
\newtheorem{thm}{Theorem}
\newtheorem{other}[lemma]{Theorem}

\def\qed{\hfill\nobreak$\Box$\break\smallskip}

\begin{document}
\maketitle
\abstract
We construct a large class of periodic instantons.  Conjecturally we
produce all periodic instantons.  This confirms a conjecture of Garland and
Murray that relates periodic instantons to orbits of the loop group acting on
an extension of its Lie algebra.\\


\section{Introduction}

Periodic instantons are solutions of the anti-self-dual equations 
\[ F_B=-*F_B\]
for a connection $B$ on a trivial vector bundle with structure group $G$ over
$S^1\times{\bf R}^3$.  In this paper, $G$ is a compact Lie group with
complexification $G^c$ equipped with a representation acting on ${\bf C}^n$
that is unitary on $G$.

Put $B=A+\Phi d\theta$ so
\begin{equation}   \label{eq:cal}
*F_A=d_A\Phi-\mu\partial_{\theta}A
\end{equation}
where we use the three-dimensional Hodge star operator and $\mu$ is
the reciprocal of the radius of the circle.  One can think of the 
connection and Higgs field as defined over ${\bf R}^3$ and dependent on the
circle-valued $\theta$.

Nahm studied periodic instantons, calling them calorons \cite{NahSel}.  Later,
Garland and Murray studied periodic instantons from the twistor viewpoint
\cite{GMuKac}.  To remedy the fact that there was so far no existence 
theorem for periodic instantons nor an understanding of the topology
of the moduli space of instantons (if they were to exist), they conjectured 
that periodic instantons can be constructed using holomorphic spheres in a 
flag manifold associated to the loop group.  This conjecture is confirmed by 
the main result of this paper, Theorem~\ref{th:instinst}.  

Recently the study of super-symmetric Yang-Mills theory over 
$S^1_{1/\mu}\times{\bf R}^3$ has been used as further evidence for the 
existence of dualities in physical theories.  In \cite{SWiGau} Seiberg and 
Witten obtained a result for periodic instantons analogous to their 1994 work
on instantons, \cite{SWiSdu}, by studying the limiting behaviour when 
$\mu\rightarrow 0$ and $\mu\rightarrow\infty$.  This led to the 
Rozansky-Witten invariants, \cite{RWiHyp}.  We will not discuss these 
developments here.

\section{Loop groups.}

Define $LG$ to be the group of smooth gauge transformations of the trivial 
$G$-bundle over the circle.  Equivalently, $LG$ is the space of smooth maps 
from $S^1$ to the compact Lie group $G$.  Following \cite{GMuKac}, intertwine
the gauge transformations with the isometries of the circle to get 
the twisted product $\widehat{LG}=LG\tilde{\times}S^1$ where the action of 
$S^1$ is given by rotation.  It has Lie algebra 
$\widehat{L\bf g}\cong L{\bf g}\oplus{\bf R}d$ with Lie bracket
\[ [X+x{\rm d},Y+y{\rm d}]=[X,Y]-y\partial X/\partial{\theta}+x\partial Y/
\partial{\theta}.\]
Put $\hat{A}=A+a{\rm d}$, $\hat{\Phi}=\Phi+\phi{\rm d}$.  Then the Bogomolny
equations over ${\bf R}^3$ for this pair are given by
\begin{equation}   \label{eq:bog} 
*F_{\hat{A}}=d_{\hat{A}}\hat{\Phi}.
\end{equation}
The d-component is given by $*da=d\phi$ so a finite energy condition will
force $a=0$ and $\phi=$constant$=\mu$, say.  The remaining part of 
(\ref{eq:bog}) is then (\ref{eq:cal}).  Thus, one can think of periodic
instantons as monopoles over ${\bf R}^3$ with structure group $\widehat{LG}$.

Monopoles for finite-dimensional groups are well-studied 
\cite{HitCon,MurMon,NahCon}.
In particular, the topology of the moduli space of monopoles is understood.
The moduli space of monopoles with structure group $G$ is diffeomorphic to
the space of holomorphic maps from the two-sphere to a homogeneous space
of $G$, or equivalently to an adjoint orbit of $G$, \cite{DonNah,JarEuc}.
In analogy with the finite-dimensional case this led Garland and Murray to 
conjecture that periodic instantons are in one-to-one correspondence with 
based holomorphic maps from $S^2$ to orbits of $\widehat{LG}$ in 
$\widehat{L\bf g}$.  The following theorem addresses half of this conjecture.
The action of $\widehat{LG}$ is really an action of $LG$.
For $(\xi,\mu)\in\widehat{L\bf g}$ denote its orbit by $LG\cdot(\xi,\mu)$.

\begin{thm}   \label{th:instinst}
There is an injective map from

(i) the space of based holomorphic maps from $S^2$ to 
$LG\cdot(\xi,\mu)$, to

(ii) the moduli space of instantons over $S^1_{1/\mu}\times{\bf R}^3$.
\end{thm}
The basing condition on the space of holomorphic maps distinguishes an
element of the orbit of $LG$ that is conjecturally the asymptotic value
of the Higgs field.  See Section~\ref{sec:bou}.  The moduli space consists
of gauge equivalence classes of connections where the gauge group consists of
gauge transformations independent of $\theta$ in the limit at infinity.

The full conjecture, that the map is also surjective, is equivalent to a
conjecture for decay properties of finite energy periodic instantons analogous
to known decay properties for monopoles.  We discuss this in 
Section~\ref{sec:bou}.

Theorem~\ref{th:instinst} can be thought of as an extension of \cite{JarMon}
from finite dimensional Lie groups to the loop group.

\subsection{Orbits of the loop group.}
The loop group $LG$ acts on $\widehat{L\bf g}$ by
\[\gamma\cdot(\xi,\mu)=(\gamma\cdot\xi-\mu\gamma'\gamma^{-1},\mu).\]
For $\xi=0$ the orbit is given by the based loop group $\Omega G$.  
More generally, we get $LG\cdot(\xi,\mu)\cong LG/Z_{\xi}$
where the isotropy subgroup $Z_{\xi}$ is described explicitly in the 
following proposition.

\begin{prop}[Pressley and Segal]
For $\pi_1G=0$ and $\mu\neq 0$ the orbits of $LG$ on $\widehat{L\bf g}$
correspond precisely to the conjugacy classes of $G$ under the map
$(\xi,\mu)\mapsto M_{\xi}\in G$ where $M_{\xi}$ is obtained by solving the
ordinary differential equation $h'h^{-1}=-\mu^{-1}\xi$ and noticing
$h(\theta+2\pi)=h(\theta)M_{\xi}$.  The isotropy subgroup of $\xi$ is given by
\begin{equation}   \label{eq:zxi}
Z_{\xi}=\{\gamma\in LG|\gamma(0)\in C[M_{\xi}], \gamma(\theta)=h(\theta)
\gamma(0)h(\theta)^{-1}\}
\end{equation}
where $C[M_{\xi}]$ is the centraliser of the conjugacy class of $M_{\xi}$ in 
$G$.
\end{prop}
Equivalently, the orbits are given by gauge equivalence classes of connections
on a trivialised bundle over the circle of radius $1/\mu$.  Each orbit is
labeled by the underlying connection which is determined by its holonomy.

In the next section we will equip the orbit of the loop group with a complex 
structure.

\subsection{Loop groups and flat connections.}   \label{sec:lgflat}
Donaldson \cite{DonBou} re-interpreted elements of the loop group in terms of
holomorphic bundles over the disk framed on the boundary, and the factorisation
theorem in terms of flat connections on these bundles.  He showed that each 
framed holomorphic bundle over the disk possesses a unique 
Hermitian-Yang-Mills (flat) connection.  

\begin{other}[Donaldson]
There is a $1-1$ correspondence between

(i) holomorphic bundles over $D$ framed over $\partial D$;

(ii) unitary Hermitian-Yang-Mills connections over $D$ on a bundle with a 
unitary framing over $\partial D$.
\end{other}

Donaldson's argument generalises to parabolic bundles---holomorphic bundles 
over the disk with a flag specified over the origin \cite{MSePar}.  In this 
case the flat connection must be singular at the origin.  

\begin{prop}   \label{th:parflat}
There is a $1-1$ correspondence between

(i) parabolic bundles over $D$ framed over $\partial D$;

(ii) unitary Hermitian-Yang-Mills connections over $D-\{0\}$ on a bundle with 
a unitary framing over $\partial D$.  The singularity of the connection at $0$
encodes the flag at $0$.
\end{prop}
Following Donaldson, we can re-interpret this result in terms of a 
factorisation theorem for loop groups as follows.  

A parabolic bundle over the disk has an underlying trivial holomorphic bundle
and a trivialisation compared to the framing over the boundary produces a loop
$\gamma\in LG^c$.  Any other trivialisation that preserves the parabolic 
structure at $0\in D$ changes $\gamma$ by an element of $L^+P$---those loops 
that are boundary values of holomorphic maps from the disk to $GL(n,{\bf C})$
with value at $0$ lying in $P$.  So (i) in the statement of 
Proposition~\ref{th:parflat} is equivalently to choosing an element of
$LG^c/L^+P$.

A unitary Hermitian-Yang-Mills (or, equivalently, flat) connection over 
$D-\{0\}$ is determined uniquely by the parabolic structure at $0\in D$.
(This would not be true if there was more than one puncture.)  With respect
to the unitary framing over the boundary, the flat connection defines
an element of the orbit $LG\cdot\xi\in\widehat{L\bf g}$.  We saw in the
previous section that the orbit is isomorphic to $LG/Z_{\xi}$.  Thus we
get the following restatement of Proposition~\ref{th:parflat}.

\begin{cor}   \label{th:fact}
For any $\xi\in L\bf g$ we have
\[ LG^c/L^+P\cong LG/Z_{\xi}.\]
\end{cor}

We could have proven the factorisation theorem in a different way.  In
the special case that $Z_{\xi}$ consists of only constant loops then 
Corollary~\ref{th:fact} follows from the standard factorisation theorem for 
loop groups.  In general, each orbit of $LG$ possesses a nice representative
which simplifies the isotropy subgroup to consist only of constant loops so 
the general case follows from the special case.

The importance of the treatment here is that at the same time as establishing
a complex structure on the orbit space, $\xi$ remains the natural base-point
for the holomorphic map and we get an interpretation of the orbit space in 
terms of flat connections over the disk on a bundle framed over the boundary.
In the next section we will see how a holomorphic map from $S^2$ into a space
of flat connections is related to an instanton over an associated 
four-manifold.

\section{Instantons and holomorphic maps into spaces of flat connections.}
Atiyah showed that there is a one-to-one correspondence between instantons 
over the four-sphere and holomorphic maps from the two-sphere to the loop 
group \cite{AtiIns}.  The interpretation of elements of the loop group in 
terms of flat connections means that Atiyah's result can be viewed as a 
relationship between instantons and holomorphic maps from the two-sphere to 
a space of flat connections.  This approach was exploited in \cite{JNoDeg}.
Another result of this type was obtained by Dostoglou and Salamon 
\cite{DosSel} in their proof of the Atiyah-Floer conjecture.  They showed 
that the instanton Floer homology associated to the three-manifold given by 
a mapping torus $S^1\tilde{\times}\Sigma$ is the same as the symplectic Floer 
homology of the space of flat connections over $\Sigma$.

The relationship between instantons and holomorphic maps into spaces of
flat connections can be understood as follows.  Suppose that locally a 
four-manifold is given by a product of two complex curves $U\times V$ 
equipped with the product metric.  The anti-self-dual equations with
respect to local coordinates $\{ w\}\times\{ z\}$ are given by:
\begin{equation}   \label{eq:asdeq}
\left.\begin{array}{l}
\lbrack\partial_{\bar{w}}^A,\partial_{\bar{z}}^A\rbrack =0\\
\lbrack\partial_{\bar{z}}^A,\partial_z^A\rbrack =\rho(w,z)\lbrack
\partial_{\bar{w}}^A,\partial_w^A\rbrack\end{array}\right\}
\end{equation}
where $\rho(w,z)$ depends on the metrics on $U$ and $V$.  

Let $f:U\rightarrow{\cal M}_V$ be a holomorphic map from $U$ into the space
of flat connections ${\cal M}_V$ over $V$.  (The conformal structure on $V$
equips the space of flat connections with a natural complex structure.)
Define a connection over $U\times V$ by
\begin{equation}  \label{eq:adff}
A=df+f(w)
\end{equation}
where $df$ is a Lie algebra valued $1$-form over $U\times V$ and $f(w)$ is 
a flat connection over $\{ w\}\times V$.  Then $A$ satisfies the following 
equations which resemble (\ref{eq:asdeq}):
\begin{equation}  \label{eq:holfl}
\left.\begin{array}{l}
\lbrack\partial_{\bar{w}}^A,\partial_{\bar{z}}^A\rbrack =0\\
\lbrack\partial_{\bar{z}}^A,\partial_z^A\rbrack =0\end{array}\right\}
\end{equation}
The first equation is equivalent to the holomorphic condition on the map $f$
and the second equation uses the fact that $f$ maps to a space of flat 
connections.

We can think of the second equation of each of (\ref{eq:asdeq}) and 
(\ref{eq:holfl}) as a type of moment map.  One can move from solutions
of (\ref{eq:holfl}) to solutions of (\ref{eq:asdeq}) using the Yang-Mills flow,
as we do in this paper or, say, by using the implicit function theorem.

In order to apply this to periodic instantons we exploit the conformal 
invariance of the anti-self-dual equations.  Let $\Sigma$ be the punctured
disk $D^2-\{0\}$ equipped with the complete hyperbolic metric 
$|dz|^2/(|z|\ln|z|)^2$.  There is a conformal equivalence:
\[ S^1\times({\bf R}^3-\{ 0\})\simeq S^2\times\Sigma,\]
where $S^1\times({\bf R}^3-\{ 0\})$ is equipped with the flat metric and
$S^2\times\Sigma$ is equipped with the product metric 
\begin{equation}   \label{eq:met} 
ds^2=\frac{4d\bar{w}dw}{(1+|w|^2)^2}+\frac{d\bar{z}dz}{|z|^2(\ln|z|)^2}.
\end{equation}
On $S^2\times\Sigma$ the anti-self-dual equations are given by 
(\ref{eq:asdeq}) with
\[\rho(w,z)=\left(\frac{1+|w|^2}{|z|\ln|z|}\right)^2.\]

Our course is set.  We have shown that a holomorphic map from $S^2$ to 
$LG\cdot(\xi,\mu)$ is the same as a holomorphic map from $S^2$ to a space of 
flat connections which gives an approximate instanton over $S^2\times\Sigma$.
In Section~\ref{sec:cons} we will use rather standard techniques to move from
an approximate instanton to an exact one.  Under the conformal equivalence 
described above, this instanton will correspond to a periodic instanton.

\subsection{Approximate instantons.}
Beginning with a holomorphic map from the two-sphere to an orbit of $LG$, we
will construct an approximate instanton over $S^1\times{\bf R}^3$.  This
will be an explicit realisation of (\ref{eq:adff}).

The map $f:S^2\rightarrow LG/Z_{\xi}$ is holomorphic when 
\[ f^{-1}\partial_{\bar{w}}f:S^2\rightarrow L^+{\bf p}\]
where $L^+{\bf p}\subset L^+{\bf g}^c$ is given by those loops that extend
to a holomorphic map of the disk whose value at the origin lies in $\bf p$.

Put $\eta$ equal to the holomorphic extension of $f^{-1}\partial_{\bar{w}}f$
to the disk.  Over $S^2\times\Sigma=\{(w,z)\}$, define a connection 
\begin{equation}   \label{eq:apcon} 
A=\eta d\bar{w}-H_{\xi}^{-1}\eta^*H_{\xi}dw+i\xi dz/z
\end{equation}
which is Hermitian with respect to the Hermitian metric 
\begin{equation}   \label{eq:inher}
H_{\xi}=exp(i\xi\ln z)^*exp(i\xi\ln z)
\end{equation}
and flat on each $\{w\}\times D$.  
Over $S^1\times{\bf R}^3$ in a radially-free gauge we get:
\[ (A,\Phi)=(\exp(i\xi r)\eta\exp(-i\xi r)d\bar{w}-\exp(-i\xi r)\eta^*
\exp(i\xi r)dw,\xi)\]
Furthermore,
\begin{equation}
*F_A=d_A\Phi-\mu\partial_{\theta}A+(1+|w|^2)^2F_{\bar{w}w}dr/r^2
\end{equation}
which resembles the periodic instanton equation, (\ref{eq:cal}).

\section{Construction}   \label{sec:cons}
In this section we will use the Yang-Mills flow to move from the 
``approximate'' periodic instanton (\ref{eq:apcon}) to an exact one.
Instead of working directly with the connections, we will follow
Donaldson \cite{DonAnt} and work with a Hermitian metric on a holomorphic
bundle which gives a Hermitian connection.  In fact, we will work with
a pair $(H,\eta)$ consisting of a Hermitian metric $H$ on a holomorphic
bundle and a map $\eta:S^2\times D^2\rightarrow{\bf g}^c$ that is
holomorphic in the second factor.  A connection $A$ is obtained from the
pair $(H,\eta)$ by:
\begin{equation}  \label{eq:metcon}
A=H^{-1}\partial_zHdz+\eta(w,z)d\bar{w}+(H^{-1}\partial_wH-H^{1}\eta(w,z)H)dw.
\end{equation}
Associate to the pair $(H,\eta)$ the Hermitian-Yang-Mills tensor
\begin{eqnarray*}  
B(H,\eta)&=&|z|^2(\ln|z|)^2\partial_{\bar{z}}(H^{-1}\partial_zH)+
(1+|w|^2)^2\{\partial_{\bar{w}}(H^{-1}\partial_wH)\\&&
-\partial_{\bar{w}}(H^{-1}
\eta^*H)-\partial_w\eta+[\eta,H^{-1}\partial_wH-H^{-1}\eta^*H]\}.
\end{eqnarray*}
When $B(H,\eta)\equiv 0$, the connection (\ref{eq:metcon}) is anti-self-dual.

Following Donaldson \cite{DonAnt} we study the heat flow for the Hermitian
metric $H$ in place of the Yang-Mills flow for the associated connection.
Since the Hermitian metrics we deal with here are not bounded we need to
extend Donaldson's results and their generalisations due to Simpson
\cite{SimCon}.  Essentially we need to understand properties of the 
Laplacian of the Kahler manifold $S^2\times\Sigma$ with metric (\ref{eq:met})
and properties of the initial Hermitian metric (\ref{eq:inher}).  Similar
results specialised to other non-compact Kahler manifolds exist in
\cite{GuoAna,JNoDeg}.

\subsection{The heat flow.}
Associate to a holomorphic map $f:S^2\rightarrow LG/Z_{\xi}$ the map
$\eta:S^2\times D^2\rightarrow{\bf g}^c$ given by the holomorphic
extension of $f^{-1}\partial_{\bar{w}}f$ to the disks in the second factor.
We would like to construct a Hermitian metric $H$ that satisfies the
equation $B(H,\eta)=0$.  This would produce an anti-self-dual connection
associated to the map $f$.

Consider the heat flow equation over $S^2\times\Sigma$
\begin{equation}   \label{eq:flow} 
H^{-1}\partial H/\partial t=B(H,\eta),\ H(w,z,0)=H_{\xi}
\end{equation}
where $H_{\xi}$ is defined in (\ref{eq:inher}).  A solution of
(\ref{eq:flow}) will converge to the required solution of $B(H,\eta)=0$ as
$t\rightarrow\infty$.  Instead of solving (\ref{eq:flow}) we will work
with a family of boundary value problems.  Put 
\[ S^2\times\Sigma_{\epsilon,\delta}=\{(w,z)\in S^2\times\Sigma\ |
\ \epsilon\leq |z|\leq\delta\}\]
so the $S^2\times\Sigma_{\epsilon,\delta}$ exhaust $S^2\times\Sigma$ as 
$\delta\rightarrow 1$ and $\epsilon\rightarrow 0$.
\begin{prop}  \label{th:ltex}
Over each $S^2\times\Sigma_{\epsilon,\delta}$ there is a unique solution 
of the boundary value problem
\begin{equation}  \label{eq:hflow}
\left.\begin{array}{c}
H^{-1}\partial H/\partial t=B(H,\eta)\\ H(w,z,0)=H_{\xi}\\
H|_{\partial S^2\times\Sigma_{\epsilon,\delta}}=H_{\xi}\end{array}\right\}
\end{equation}
given by $H^{\epsilon,\delta}(w,z,t)$ and converging to a smooth metric 
$H^{\epsilon,\delta}(w,z,\infty)$ that satisfies 
$B(H^{\epsilon,\delta}(w,z,\infty),\eta)=0$.
\end{prop}
{\em Proof.}  Since we have fixed $S^2\times\Sigma_{\epsilon,\delta}$ for the 
moment we will omit the superscript in $H^{\epsilon,\delta}(w,z,t)$ 
during this proof.  Short-time existence of a solution of (\ref{eq:hflow}) is 
automatic since $B(H,\eta)$ is elliptic in $H$ and we have Dirichlet
boundary conditions.  In order to extend this to long-time existence
we will take the approach given by Donaldson \cite{DonAnt} and extended
by Simpson \cite{SimCon} and show that a solution on $[0,T)$
gives a limit at $T$ which is a good initial condition to start the flow
again.  The lemmas we need to prove on the way use the details of
our particular case and allow us to proceed with Donaldson's proof.

A Hermitian metric $H$ takes its values in the space $G^c/G$
which comes equipped with the complete metric $d$ given locally by
$tr(H^{-1}\delta H)^2$.  Following Donaldson, we will use both this
metric and the convenient function 
$\sigma(H_1,H_2)=tr(H_1^{-1}H_2)+tr(H_1H_2^{-1})-2n$ 
that satisfies $c_1d^2\leq\sigma\leq c_2d^2$ for constants $c_1,c_2$.
(Aside:  if we take the loop group perspective described in \cite{GMuKac},
then a Hermitian metric takes its values in the space $LG^c/LG$.  We
have not checked that this is a complete metric space.)

\begin{lemma}
If $H_1$ and $H_2$ are two solutions of the heat equation then
\begin{equation}   \label{eq:diffc}
\partial_t\sigma+\Delta\sigma\leq 0
\end{equation}
for $\sigma=\sigma(H_1,H_2)$.  
\end{lemma}
{\em Proof.} See \cite{JNoDeg}.\qed

Apply (\ref{eq:diffc}) to $H(w,z,t)$ and $H(w,z,t+\tau)$, the flow at two 
times.  Since they obey the same boundary conditions on 
$S^2\times\Sigma_{\epsilon,\delta}$, $\sigma$ vanishes on the boundary.  
By the maximum principle $\sup_{S^2\times\Sigma_{\epsilon,\delta}}\sigma$ 
is a non-increasing function of $t$.  By continuity, for any $\rho>0$ there 
exists a $\tau$ small enough so that
\[\sup_{S^2\times\Sigma_{\epsilon,\delta}}\sigma(H(w,z,t),H(w,z,t'))<\rho\]
for $0<t,t'<\tau$.  It follows from the non-increasing property of $\sigma$
that
\[\sup_{S^2\times\Sigma_{\epsilon,\delta}}\sigma(H(w,z,t),H(w,z,t'))<\rho\]
for $T-\tau<t,t'<T$.  Since $\rho$ can be made arbitrarily small,
$H(w,z,t)$ is a Cauchy sequence in the $C^0$ norm as $t\rightarrow T$.  The
metrics take their values in a complete metric space (described below)
and the function $\sigma$ acts like the metric so there is a {\em continuous}
limit $H_T$ of the sequence.  Notice also that (\ref{eq:diffc}) and the
maximum principle show that this short-time solution to the heat flow
equation is unique.

Using the heat equation and the metric on $G^c/G$, we have
\[ d(H(w,z,t),H(w,z,0))\leq\int^t_0|B(H(w,z,s),\eta)|ds\]
where $|B(H(w,z,s),\eta)|^2=tr(B^*B)$ and the adjoint is taken with respect to 
the metric $H_s$.  Notice that $B^*=B$ so $|B(H(w,z,s),\eta)|^2=tr(B^2)$.

\begin{lemma}   \label{th:bin}
If $H(w,z,t)$ is a solution of the heat equation then
\begin{equation}  \label{eq:bin}
(d/dt+\Delta)|B(H(w,z,t),\eta)|\leq 0\ {\rm whenever}\ |B|>0
\end{equation}
\end{lemma}
{\em Proof.} See \cite{JNoDeg}.\qed

The next two lemmas use the particular features of the Kahler manifold
$S^2\times\Sigma$ together with the initial Hermitian metric $H_{\xi}$ to
get $C^0$ control on $H(w,z,t)$ during the flow.

\begin{lemma}  \label{th:initb}
When $\eta$ is the holomorphic extension of $f^{-1}\partial_{\bar{w}}f$, 
for a given holomorphic map $f:S^2\rightarrow\Omega LG/Z_{\xi}$, there exists
a constant $M$ such that $|B(H_{\xi},\eta)|\leq M(1-|z|)$ on $S^2\times\Sigma$.
\end{lemma}
{\em Proof.}  
\[ B(H_{\xi},\eta)=-(1+|w|^2)^2(\partial_w\eta+\partial_{\bar{w}}
(H_{\xi}^{-1}\eta^*H_{\xi})+[\eta,H_{\xi}^{-1}\eta^*H_{\xi}])\]
and since $[\eta(0),\xi]=0$, $|B(H_{\xi},\eta)|$ is bounded near $z=0$.
Since $f$ takes its values in the unitary loop group and $H_{\xi}=I$ on 
$|z|=1$, we can identify $B(H_{\xi},\eta)$ with the curvature of a flat 
connection which is $0$.  Furthermore, $B(H_{\xi},\eta)$ is continuous and
differentiable up to $|z|=1$ so it vanishes like $1-|z|$ there.  \qed

\begin{lemma}  \label{th:disbt}
There is a constant $C$ independent of $\epsilon$ and $\delta$ such that
\[d(H^{\epsilon,\delta}(w,z,t),H_{\xi})\leq C\ln(1-\ln |z|)\]
for all $(w,z,t)\in S^2\times\Sigma_{\epsilon,\delta}\times\bf R$.
\end{lemma}
{\em Proof.}  It follows from (\ref{eq:bin}) and the maximum principle that 
if there is a function $b(w,z,t)$ defined on 
$S^2\times\Sigma_{\epsilon,\delta}\times{\bf R}$ that satisfies 
$(\partial_t+\Delta)b=0$ and $|B(H_{\xi},\eta)|\leq b(w,z,0)$
then $|B(H(w,z,t),\eta)|\leq b(w,z,t)$ for all $t$.

Put $b(w,z,0)=M(1-|z|)$.  Notice that $b(w,z,0)=b(|z|)$, so we only need use
the one-dimensional Laplacian and $b(w,z,t)=b(|z|,t)$.  From the flow 
equation (\ref{eq:hflow}) we have
\begin{eqnarray}
d(H(w,z,t),H_{\xi}(w,z))&=&\int^t_0B(H(w,z,\tau))d\tau\nonumber\\
&\leq&\int_0^tb(w,z,\tau)d\tau\nonumber\\
&\leq&\int_0^{\infty}b(w,z,\tau)d\tau  \label{eq:db}
\end{eqnarray}
Now, $b(|z|,t)=\int b(s,t)k(|z|,s,t)ds$ where $k$ is the 
one-dimensional heat kernel operator.  Since 
$\int^{\infty}_0k(|z|,s,t)dt=G(|z|,s)$, the Green's operator, is finite, 
Fubini's theorem allows us to interchange the order of integration in 
(\ref{eq:db}).  So 
\begin{eqnarray*}
d(H(w,z,t),H_{\xi}(w,z))&\leq&M\int^{\epsilon}_0(1-s)G(|z|,s)ds\\
&\leq&M\int^1_0(1-s)G(|z|,s)ds\ .
\end{eqnarray*}
With respect to the Laplacian 
\[\Delta=-(1+|w|)^2\partial_{\bar{w}}\partial_w-4|z|^2(\ln|z|)^2\partial
_{\bar{z}}\partial_z=-(\ln|z|)^2\partial^2_{\ln|z|}\]
reduced to one dimension, the Green's operator is given by
\[ G(|z|,s)={\rm min}\{ -\ln |z|,-\ln s\}/s(\ln s)^2\ .\] 
Actually, this Green's operator is only valid for the
entire interval ($\epsilon=1$) and Fubini's theorem doesn't apply there.
There is a monotone property of heat kernels which means that our
choice of $G$ is simply an overestimate when $\epsilon<1$ so the calculation
is valid.  Thus 
\begin{eqnarray*}    
d(H(w,z,t),H_{\xi}(w,z))&\leq&M\left(-\ln |z|\int^{|z|}_0\frac{(1-s)ds}
{s(\ln s)^2}-\int^1_{|z|}\frac{(1-s)ds}{s\ln s}\right)\\
&\leq&C\ln(1-\ln |z|)  
\end{eqnarray*}
where the last inequality simply encodes the fact that the distance
vanishes as $|z|\rightarrow 1$ and grows like $\ln(1-\ln |z|)$ as
$|z|\rightarrow 0$.  \qed

The preceding lemmas have shown that there is a solution to the heat equation
that satisfies $H(w,z,t)\rightarrow H(w,z,T)$ in $C^0$ and $H(w,z,t)$ is 
uniformly bounded with bound independent of $t$ (though depending on 
$\epsilon$).  These are the conditions required to use Simpson's extension of 
Donaldson's result to show that $H(w,z,t)$ is bounded in $L^p_2$ uniformly in 
$t$.  Hamilton's methods \cite{HamHar} then give control of all higher Sobolev
norms.  Thus we get a solution, $H(w,z,t)$, of (\ref{eq:hflow}) for all $t$ 
that converges to a smooth limit $H^{\epsilon,\delta}(w,z,\infty)$ defined on 
$S^2\times\Sigma_{\epsilon,\delta}$ and satisfying 
$B(H^{\epsilon,\delta}(w,z,\infty),\eta)=0$ and 
$H^{\epsilon,\delta}(w,z,\infty)=H_{\xi}$ on 
$\partial S^2\times\Sigma_{\epsilon,\delta}$ so Proposition~\ref{th:ltex} 
is proven.   \qed

\begin{prop}  \label{th:exist}
For each holomorphic map $f:S^2\rightarrow LG/Z_{\xi}$ there is a
periodic instanton $A_f$ on $S^1\times{\bf R}^3$.
\end{prop}
{\em Proof.}  We have proven the existence of a family of hermitian metrics
$H^{\epsilon,\delta}$ respectively defined over 
$S^2\times\Sigma_{\epsilon,\delta}$ and satisfying 
$B(H^{\epsilon,\delta},\eta)=0$.  Since 
$\sigma(H^{\epsilon,\delta},H^{\epsilon',\delta'})$ is subharmonic its 
maximum occurs at the boundary of the set on which it is defined.  For 
$\epsilon'\leq\epsilon\leq\delta\leq\delta'$, the common set is 
$S^2\times\Sigma_{\epsilon,\delta}$.  If we fix $\epsilon=\epsilon'$ and let
$\delta\rightarrow 1$, then $\sigma=0$ on $|z|=\epsilon$ and the maximum
of $\sigma$ occurs on $|z|=\delta$.  Since the metrics $\sigma$ and $d$ on
$G^c/G$ are equivalent, the maximum value of $\sigma$ is bounded by a
constant times $d(H^{\epsilon,\delta'},H_{\xi})\leq C\ln(1-\ln\delta)$ using
Lemma~\ref{th:disbt}.  This tends to $0$ as $\delta\rightarrow 1$, thus we 
have a Cauchy sequence that converges uniformly to a Hermitian metric 
$H^{\epsilon}$ defined on $|z|\geq\epsilon$.  The convergence can be improved 
to $L^p_2$ to ensure that $B(H^{\epsilon},\eta)=0$, \cite{SimCon}.

In order to deal with $\epsilon\rightarrow 0$, notice that since $\ln|z|$
is harmonic on $S^2\times\Sigma$, $\sigma+a\ln|z|$ is subharmonic for any
$a$.  Put $a=\sup_{|z|=\epsilon}\sigma/|\ln\epsilon|$.  Then 
$\sigma+a\ln|z|\leq 0$ on $|z|=1$ and $|z|=\epsilon$.  Thus
\begin{equation}   \label{eq:cauchy}
\sigma\leq -\ln|z|\sup_{|z|=\epsilon}\sigma/|\ln\epsilon|.
\end{equation}
By Lemma~\ref{th:disbt}, 
$d(H^{\epsilon,\delta'},H_{\xi})\leq C\ln(1-\ln\epsilon)$
so $\sigma=o(|\ln\epsilon|)$ as $\epsilon\rightarrow 0$.  Thus the right hand
side of (\ref{eq:cauchy}) tends uniformly to $0$ on compact sets away from
$z=0$.  Again we conclude that the $\{ H^{\epsilon}\}$ form a Cauchy sequence
as $\epsilon\rightarrow 0$, converging uniformly on the complement of any
neighbourhood of $S^2\times\{ 0\}$ to a Hermitian metric $H$ that satisfies
$B(H,\eta)=0$ on $S^2\times\Sigma$.

Using $S^1\times({\bf R}^3-\{ 0\})\cong S^2\times\Sigma$ we see that the
limit $H$ is smooth on $S^1\times({\bf R}^3-\{ 0\})$ and continuous on all
of $S^1\times{\bf R}^3$, converging to $I$ on $S^1\times\{ 0\}$.
The connection $A$ obtained from $H$ via (\ref{eq:metcon}) is defined
and anti-self-dual on $S^1\times({\bf R}^3-\{ 0\})$.  By the following 
lemma, $A$ has finite charge.  Since codimension three singularities of 
finite charge anti-self-dual connections can be removed \cite{SSiCla}, 
$A$ is smooth on all of $S^1\times{\bf R}^3$.  \qed

\begin{lemma}  \label{th:finch}
The curvature of the limiting connection $A$ has finite $L^2$ norm.
\end{lemma}
{\em Proof.}  The Yang-Mills flow decreases the $L^2$ norm of a connection,
and any bubbling in the limit just decreases the $L^2$ norm further,
so it is sufficent to show that the initial connection has finite $L^2$ norm.

For any connection $A$, we have
\begin{equation}   \label{eq:l2norm}
8\pi^2\|F_A\|_2^2=2\int |F_A^+|^2-\int F_A\wedge F_A
\end{equation}
where $F_A^+$ is the self-dual part of the curvature.  We can calculate this
explicitly for the initial connection defined in (\ref{eq:apcon}).  

Notice that $F_A^+=B(H_{\xi},\eta)$ and by Lemma~\ref{th:initb} we have
$|B(H_{\xi},\eta)|\leq M(1-|z|)$.  This is square-integrable over 
$S^2\times\Sigma$ since $S^2$ is compact and $\Sigma$ has finite area near 
$z=0$ and grows like $1/(1-|z|)^2$ near $|z|=1$.

As one might expect, the topological term in (\ref{eq:l2norm}) will
coincide with the topological degree of the map $f:S^2\rightarrow LG/Z_{\xi}$.
\[ k(E)=\frac{1}{8\pi^2}\int_{S^2\times D}tr(F_A^2)=-\frac{1}{8\pi^2}\int_{S^2
\times D}tr(\partial_{\bar{z}}\eta^*\partial_z\eta)d\bar{z}dzd\bar{w}dw\]
since only the $F_{z\bar{w}}$ and $F_{\bar{z}w}$ terms contribute.  Since
$\eta$ is holomorphic in $z$, then on the disk 
$d\{ tr(\eta^*\partial_z\eta) dz\}=
tr(\partial_{\bar{z}}\eta^*\partial_z\eta)d\bar{z}dz$ so 
\begin{eqnarray*}
k(E)&=&-\frac{1}{8\pi^2}\int_{S^2}\int_{|z|=1}tr(\eta^*\partial_z\eta)
dzd\bar{w}dw\\
&=&\frac{1}{4\pi}\int_{S^2}\|f^{-1}\partial_{\bar{w}}f\|^2\frac{d\bar{w}dw}{i}
\end{eqnarray*}
where $\|f^{-1}\partial_{\bar{w}f}\|^2$ uses the Kahler metric on $LG/Z_{\xi}$.
This expression is the degree of $f$.  \qed

{\em Remark.}  In the construction of this section we started with parabolic
bundles over the disk.  However, the reverse is not true that a periodic 
instanton gives rise to a family of parabolic bundles.  By this we mean that 
the holomorphic structure defined on each punctured disk by the restriction
of the periodic instanton does not extend to the entire disk.  The curvature 
just fails to satisfy $F_A\in L^p$ for $p>1$ as required in \cite{BiqPar}.

\section{Injection}   \label{sec:inj}

In this section we will show that the map produced in Section~\ref{sec:cons} 
is injective.
\begin{prop}
Let $f:S^2\rightarrow LG/Z_{\xi}$ and $g:S^2\rightarrow LG/Z_{\nu}$ be two
based holomorphic maps.  Then the instantons $A_f$ and $A_g$ are gauge
equivalent precisely when $\nu-\xi$ is in the root lattice and 
$g=f\cdot\exp(i(\nu-\xi)\ln z)$.
\end{prop}
{\em Proof.}  The instanton $A_f$ is given by the expression (\ref{eq:metcon})
which depends on a pair $(H,\eta)$ consisting of a Hermitian metric, $H$, and 
the holomorphic extension of $f^{-1}\partial_{\bar{w}}f$ denoted by $\eta$
and likewise for $A_g$.  These expressions are independent of the unitary
gauge so $A_f\sim A_g$ only if $A_f=A_g$ or possibly if we have used different
holomorphic trivialisations of the holomorphically trivial bundle restricted
to each $\{ w\}\times\Sigma$ for $A_f$ and $A_g$.

If $A_f=A_g$ then $f^{-1}\partial_{\bar{w}}f=\eta=g^{-1}\partial_{\bar{w}}g$,
so $\partial_{\bar{w}}(gf^{-1})=0$ and this is global over $S^2$ thus
$g=\gamma(z)f$ for some loop $\gamma(z)$ independent of $w$.  The requirement
that $f$ and $g$ map $\infty\in S^2$ to the constant loop $I$ forces 
$\gamma(z)=I$.  

If $A_f\neq A_g$ and $A_f\sim A_g$ then $A_f$ uses the pair $(H,\eta)$ in
(\ref{eq:metcon}) and $A_g$ uses the pair
$(p^*Hp,p^{-1}\eta p+p^{-1}\partial_{\bar{w}}p)$ for a map 
$p:S^2\times\Sigma\rightarrow G^c$ which is holomorphic on each 
$\{ w\}\times\Sigma$ and unitary on its boundary.  Note that this implies that
$g=fp$ though since $p$ is not a priori in $L^+P$, the maps $f$ and $g$ can 
be distinct.

The proof of the proposition is completed by the following two 
lemmas that show that $g=fp$ together with the known growth of the
Hermitian metrics associated to $f$ and $g$ forces $p$ to be constant
or to be a standard holomorphic gauge change.

\begin{lemma}
If $\xi=\nu$ then $A_f\sim A_g$ only if $f=gu$ for $u\in P\cap G\cong Z_{\xi}$.
\end{lemma}
{\em Proof.}  We can apply Lemma~\ref{th:disbt} to the Hermitian-Yang Mills 
metric $H$ over all of $S^2\times\Sigma$ even though it is only stated
for $0<\epsilon<\delta<1$.  Thus
\[ d(H,H_{\xi})+d(p^*Hp,H_{\xi})\leq C\ln(1-\ln|z|)\]
for the initial metric $H_{\xi}$ defined in (\ref{eq:inher}).  Using the
identity $d(p^*Hp,H_{\xi})=d((p^*)^{-1}H_{\xi}p^{-1},H_{\xi}))$ and the 
triangle inequality we have
\begin{equation}    \label{eq:ineq}
d(H,H_{\xi})+d(p^*Hp,H_{\xi})\geq d((p^*)^{-1}H_{\xi}p^{-1},H_{\xi}))
\end{equation}
and the right hand side is bounded by $C\ln(1-\ln|z|)$ only if $p$ is 
bounded near $z=0$ by $C\ln(1-\ln|z|)$.  Since it satisfies 
$\lim_{z\rightarrow 0}zp(z)\rightarrow 0$, $p$ extends across $z=0$ and is 
holomorphic there.  Furthermore we must have $p(0)\in P$ in order that the 
right hand side of (\ref{eq:ineq}) is bounded by $C\ln(1-\ln|z|)$.  Since $p$ 
is holomorphic on the disk and unitary on the boundary it must be unitary
on the disk (by the maximum principle applied to the subharmonic function
$tr(p^*p)+tr((p^*p)^{-1})$), and thus constant there, and moreover lie in
$P\cap G$.  \hfill\nobreak$\Box$
\begin{lemma}
If $A_f\sim A_g$ then $\nu-\xi$ lies in the root lattice and 
\[ g=f\exp(i(\nu-\xi)\ln z).\]
\end{lemma}
{\em Proof.}  As described above, $g=fp$.  Then
$\lim_{z\rightarrow 0}zp^{-1}\partial_zp=\nu-\xi$.  Since $zp^{-1}\partial_zp$
is bounded and holomorphic on the punctured disk, it extends to a holomorphic
function of the disk.  In fact $p^{-1}\partial_zp=q(z)/z$ so 
$p(z)=\exp(\int^zq(\zeta)d\zeta/\zeta)$ and $\nu-\xi=q(0)$ must lie in the
integer lattice.  Thus $p\cdot\exp(-i(\nu-\xi)\ln z)$ is holomorphic on the 
disk and unitary on the boundary and hence constant which we absorb in the 
unitary ambiguity of $f$.  So $g=f\cdot\exp(i(\nu-\xi)\ln z)$.  \qed

The proposition allowed for gauge transformations that have angular dependence
at infinity (corresponding to $z=0$).  When we restrict the gauge 
transformations to have no angular dependence at infinity then the maps
$f$ and $f\cdot\exp(i(\nu-\xi)\ln z)$ define inequivalent connections.
Thus the map $f\mapsto A_f$ is injective.

\section{Boundary conditions}    \label{sec:bou}
There are natural boundary conditions that the periodic instantons constructed
in this paper conjecturally satisfy:  as $r\rightarrow\infty$
\[\begin{array}{c} \|\Phi-\xi\|=O(1/r)\\ \partial\|\Phi-\xi\|/\partial
\Omega=O(1/r^2)\\ \|\nabla(\Phi-\xi)\|=O(1/r^2)
\end{array}\]
where $\xi$ is a given constant Higgs field, $r$ is the radial coordinate 
in ${\bf R}^3$, $\partial/\partial\Omega$ is an angular derivative, and the 
asymptotic constants are uniform in $\theta$.

In order to prove these conditions we would need to understand the precise
elliptic constants for the Hermitian Yang-Mills Laplacian on $S^2\times\Sigma$
near the puncture at $z=0$.  This would enable us to get estimates on
the second derivatives of $H$ from the estimates on $H$ given in this paper
and estimates on first derivatives of $H$ obtained from a maximum principle
argument \cite{DonBou}.  We hope to show this in future work.

Alternatively, one might prove the stronger conjecture that all finite energy
periodic instantons satisfy these boundary conditions.  Such a proof would
again require a good understanding of the Laplacian on $S^2\times\Sigma$
as in the special case of monopoles \cite{JTaVor}.  This stronger conjecture
implies that the construction of this paper yields {\em all} periodic 
instantons.  This can be proven by using a scattering argument to retrieve
a holomorphic map from $S^2$ to an orbit of the loop group from a given 
periodic instanton.  \\
\\

{\em Acknowledgements.}  I would like to thank Michael Murray for useful
discussions and the University of Adelaide for its hospitality over a
period when part of this work was carried out.

\vskip .1in
\noindent
{\small 

Paul Norbury

Department of Mathematics and Statistics, 

University of Melbourne, 

Victoria, 3052,

Australia 

norbs@ms.unimelb.edu.au}

\end{document}